\documentclass[sigconf]{acmart}
\usepackage{xspace}
\usepackage[table]{xcolor}   
\usepackage{booktabs}
\usepackage{multirow}
\usepackage{CJKutf8}      
\usepackage{tabularx}
\usepackage{array}
\usepackage{makecell}   
\usepackage{ragged2e}
\usepackage{enumitem}
\definecolor{headblue}{RGB}{40,70,130}
\definecolor{rowgray}{RGB}{240,242,247}
\newcolumntype{C}{>{\centering\arraybackslash}m{2.6cm}}
\newcolumntype{L}{>{\RaggedRight\arraybackslash}X}   

\newcommand{\eg}{\textit{e.g.}\xspace}

\AtBeginDocument{%
  }

\setcopyright{acmlicensed}
\copyrightyear{2018}
\acmYear{2018}
\acmDOI{XXXXXXX.XXXXXXX}
\acmConference[Conference acronym 'XX]{Make sure to enter the correct
  conference title from your rights confirmation email}{June 03--05,
  2018}{Woodstock, NY}
\acmISBN{978-1-4503-XXXX-X/2018/06}

\begin{document}

\title{SLIDE: Search-to-List Item Distillation for Single-Rollout Generative Retrieval}


\author{Tianyu Zhan}
\authornote{Equal contribution. Work is done during Tianyu Zhan's internship at Alibaba Group.}
\affiliation{%
  \institution{Zhejiang University}
  \city{Hangzhou}
  \country{China}}
\email{yuzt@zju.edu.cn}

\author{Gui Ling}
\authornotemark[1]
\affiliation{%
  \institution{Taobao \& Tmall Group of Alibaba}
  \city{Hangzhou}
  \country{China}}
\email{linggui.lg@taobao.com}

\author{Ran Tao}
\affiliation{%
  \institution{Taobao \& Tmall Group of Alibaba}
  \city{Hangzhou}
  \country{China}}
\email{linkunhai.lkh@taobao.com}

\author{Kunhai Lin}
\affiliation{%
  \institution{Taobao \& Tmall Group of Alibaba}
  \city{Hangzhou}
  \country{China}}
\email{linkunhai.lkh@taobao.com}

\author{Tong Xiong}
\affiliation{%
  \institution{Taobao \& Tmall Group of Alibaba}
  \city{Hangzhou}
  \country{China}}
\email{xiongtong.xt@taobao.com}

\author{Yang Wang}
\affiliation{%
  \institution{Taobao \& Tmall Group of Alibaba}
  \city{Hangzhou}
  \country{China}}
\email{wy416570@taobao.com}

\author{Kaixuan Zhang}
\affiliation{%
  \institution{Taobao \& Tmall Group of Alibaba}
  \city{Hangzhou}
  \country{China}}
\email{zhangkaixuan.zkx@taobao.com}

\author{Zhihong Chen}
\authornote{Corresponding author.}
\affiliation{%
  \institution{Taobao \& Tmall Group of Alibaba}
  \city{Hangzhou}
  \country{China}}
\email{jhon.czh@taobao.com}

\author{Yuliang Yan}
\affiliation{%
  \institution{Taobao \& Tmall Group of Alibaba}
  \city{Hangzhou}
  \country{China}}
\email{yuliang.yyl@taobao.com}

\author{Dan Ou}
\affiliation{%
  \institution{Taobao \& Tmall Group of Alibaba}
  \city{Hangzhou}
  \country{China}}
\email{oudan.od@taobao.com}

\author{Shengyu Zhang}
\authornotemark[2]
\affiliation{%
  \institution{Zhejiang University}
  \city{Hangzhou}
  \country{China}}
\email{sy_zhang@zju.edu.cn}

\author{Haihong Tang}
\affiliation{%
  \institution{Taobao \& Tmall Group of Alibaba}
  \city{Hangzhou}
  \country{China}}
\email{piaoxue@taobao.com}


\renewcommand{\shortauthors}{Trovato et al.}

\begin{abstract}
Generative retrieval (GR) typically relies on item-level beam search to retrieve candidates, leaving cross-item dependencies unmodeled and limiting direct optimization of list-level objectives. A natural alternative is to generate the entire list within a single autoregressive trajectory. However, greedy decoding with directly trained list-wise models substantially underperforms beam search. We identify two contributing factors: the broader search space explored by beam search and the positional bias of scarce positive signals, which deprives later positions of adequate supervision. To address these challenges, we propose \textbf{SLIDE}, a search-to-list item distillation framework that decouples offline search from online list generation. A teacher model retrieves high-quality candidates through offline item-level beam search, providing additional conversion-relevant supervision beyond sparse behavioral labels. Their Semantic IDs (SIDs) are then concatenated in beam-score order to form list-wise distillation targets. This \emph{search-to-list conversion} transforms candidates retrieved along separate search paths into sequential item-generation targets within a single trajectory. Trained on these targets, the student generates the Top-$K$ list in a single greedy pass, also decoupling teacher scaling from online deployment. Experiments on FORGE and Taobao Search show that SLIDE substantially improves retrieval quality under single greedy decoding and, with stronger offline teachers, surpasses the beam-search reference. 
Supplementary RL experiments support the effectiveness of list-level generation.

\end{abstract}

\ccsdesc[500]{Information systems~Retrieval models and ranking}

\keywords{Generative Retrieval, Search-to-List Distillation, List-wise Autoregressive Generation}


\maketitle

\section{Introduction}
\label{sec:intro}

Generative retrieval (GR) reframes candidate retrieval as autoregressive sequence generation~\cite{dsi,tiger}, assigning each item fixed-length Semantic IDs (SIDs) and training a single model to generate SIDs directly~\cite{forge,gpr}. This formulation mitigates item cold-start challenges and avoids unbounded vocabulary growth, supporting its increasing adoption in industrial search, advertising, and recommendation systems~\cite{onesearch-v1,onesearch-v2,univa,onerec_report,gr4ad,CRSID}.

Despite these advances, the dominant GR paradigm combines \emph{point-wise} training with \emph{item-level beam search} at inference~\cite{onesearch-v1,onerec_report,onesug,oneloc,tsgr,forge,gpr}: each target is supervised independently, and beam search produces the Top-$K$ list. This combination has two limitations:
\begin{itemize}[leftmargin=*, nosep]
\item \textbf{No explicit cross-item modeling.} Candidates are generated independently under a shared request context, without explicitly conditioning on previously retrieved items.

\item \textbf{Limited alignment with list-level objectives.} By making candidates compete for the top-1 position, item-level reinforcement learning (RL) struggles to optimize list-level objectives such as diversity~\cite{unisearch} and may compromise deeper coverage~\cite{tsgr,CQSID,genrec_netflix}.

\end{itemize}

\begin{figure}[t]
    \centering
    \includegraphics[width=0.98\linewidth]{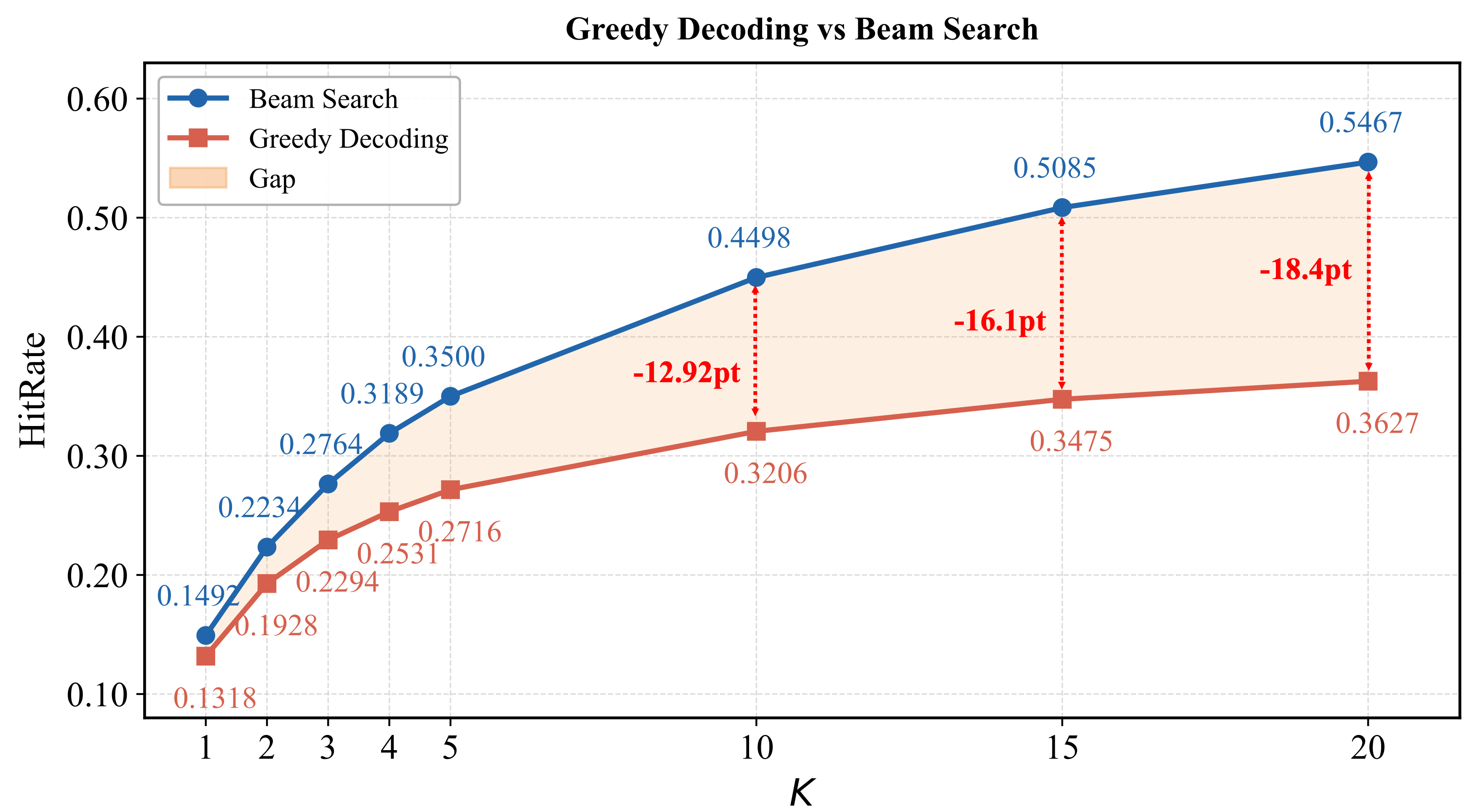}
    \caption{Greedy decoding vs.\ beam search using the same directly trained list-wise model on Taobao Search.}
    \label{fig:intro}
\end{figure}

A natural alternative is \emph{list-wise single greedy decoding}~\cite{unisearch}, which generates the entire list in a single autoregressive trajectory, with each item conditioned on the preceding items. However, when both decoding strategies use the same directly trained list-wise model, greedy decoding underperforms beam search by $18.4\%$ in HR@20 on Taobao Search (Figure~\ref{fig:intro}). Two factors contribute to this performance gap:
\begin{itemize}[leftmargin=*, nosep]
\item \textbf{Larger search space.} Beam search explores more SID sequences than greedy decoding, enabling it to identify items with high joint SID probabilities. The advantage is evident at HR@1.
\item \textbf{Scarce positive supervision.}
In interaction-ordered targets, converted items precede clicked and impressed items. Because conversions are sparse, list position can become associated with interaction type, weakening supervision for retrieving relevant items across all list positions.
This supervision bias contributes to the widening HR gap at larger cutoffs.
\end{itemize}

The strong performance of beam search suggests that the trained model has learned relevance patterns that single greedy decoding does not fully exploit. From a distributional perspective, existing relevance supervision provides only a partial view of the retrieval space: behavioral labels reflect interactions with exposed items, while fine-ranking targets cover candidates selected by upstream retrieval. Direct training therefore learns from a distribution shaped by the existing system's candidate-selection and exposure mechanisms, although retrieval must identify relevant items across the full item catalog. This raises a natural question: can we reconstruct supervision beyond the observed candidate subset and make it accessible to single greedy decoding?
Our key insight is to replace supervision restricted to observed candidates with list-wise targets derived from a teacher-induced retrieval distribution over the full item catalog. The teacher uses relevance patterns learned from existing supervision to identify promising candidates beyond those observed for each request. Crucially, broader candidate coverage alone does not teach a greedy decoder how to recover these candidates within one trajectory. The supervision must also organize them into an ordered list, enabling the student to learn successive retrieval decisions conditioned on preceding items.
We instantiate this idea as \textbf{SLIDE} (\textit{\textbf{S}earch-to-\textbf{L}ist \textbf{I}tem \textbf{D}istillation for singl\textbf{E}-rollout generative retrieval}). Its \emph{search-to-list conversion} turns offline item-level beam-search outputs into list-wise distillation targets, transferring both candidate coverage and ordering into a single autoregressive trajectory. Rather than requiring online search to recover candidates from multiple branches, SLIDE trains the student to generate the full Top-$K$ list in a single greedy pass. This decouples multi-paths search from online serving~\cite{degre}, allowing teacher capacity to scale without changing the deployed model or its online serving requirements.



This work presents an approach to improving list-wise GR under single greedy decoding. Our main contributions are summarized as follows:
\begin{itemize}[leftmargin=*, nosep]
    \item We characterize the performance gap between item-level beam search and single greedy decoding, identifying broader search exploration and sparse conversion supervision in observed behavioral data as two contributing factors.

    \item We propose \textbf{SLIDE}, a search-to-list distillation framework that transfers offline item-level search into list-wise generation, enabling full Top-$K$ retrieval through single greedy decoding while decoupling teacher scaling from online deployment.

    \item Extensive experiments on FORGE~\cite{forge} and in-house Taobao Search show that SLIDE substantially improves greedy retrieval and, with stronger offline teachers, surpasses the beam-search reference at the same deployment size. Our analyses reveal how supervision design and scaling affect this transfer, while supplementary RL experiments support the effectiveness of list-level generation for subsequent policy refinement.
\end{itemize}

\section{Related Work}
\subsection{Generative Retrieval and Recommendation}

Generative retrieval and recommendation formulate item retrieval as sequence generation~\cite{dsi,tiger}, a paradigm increasingly adopted in industrial search, advertising, and recommendation systems~\cite{onesearch-v2,univa,onerec_report,gr4ad,zhan2026semantic}. Many existing methods combine point-wise training with item-level beam search to retrieve candidate lists, sometimes followed by reranking~\cite{onesearch-v1,onerec_report,onesug,oneloc,tsgr,forge,gpr}. Item-level beam search retrieves candidates along separate SID paths under a shared context rather than making sequential item-generation decisions conditioned on preceding items. Consequently, it does not explicitly model cross-item dependencies. RL methods such as OneSearch-V2~\cite{onesearch-v2} and UniVA~\cite{univa} also optimize at the item level: each rollout generates one item and receives item-level rewards, limiting direct optimization of list-level objectives. UniSearch~\cite{unisearch} identifies point-wise beam generation as a limitation for result diversity and proposes list-wise generation as a direction for future work.

Another line of work uses sessions or pages as the unit of generation to capture cross-item dependencies~\cite{nlgr}. OneRec-V1~\cite{onerec-v1} models cross-video dependencies through session-wise learning and a session-level reward model. EGA-V2~\cite{ega-v2} learns high-value ad sequences from logged impressions and uses a reward model to select among beam-search candidates. GenRec~\cite{genrec} uses page-wise supervision during training but returns to item-level beam search at inference, separating list-wise training from item-wise retrieval. GenPage~\cite{genpage} serializes structured multi-row pages and explores page-level RL. These works demonstrate the value of list-wise generation for cross-item modeling and list-level optimization. However, most still rely on the exploratory capacity of beam search at inference. SLIDE instead enable the model to generate the full list in a single greedy pass.

\subsection{Knowledge Distillation}
Knowledge distillation transfers knowledge from a teacher to a student, often reducing serving costs while retaining prediction quality~\cite{rankdistill,rankgpt,gryphon}. Beyond model capabilities, distillation can also transfer the benefits of computationally expensive decoding. DeGRe~\cite{degre} applies offline exploration and distillation to generative reranking over a given candidate set. It trains an online generator with sequence-level hard targets and step-wise soft supervision, enabling list generation through single greedy decoding.
Whereas DeGRe preserves the list-wise output structure of the reranking task, SLIDE converts item-level beam-search outputs into list-wise generation targets for retrieval. The teacher retrieves candidates through item-level beam search, and their SIDs are concatenated in beam order into a single distillation target. This \emph{search-to-list conversion} transforms candidates retrieved along separate search paths into consecutive item-generation decisions within one autoregressive trajectory. The resulting targets provide ordered supervision across list positions, reducing reliance on scarce conversion labels and supporting cross-item conditional learning.

\section{Preliminaries}
\begin{figure}
    \centering
    \includegraphics[width=0.98\linewidth]{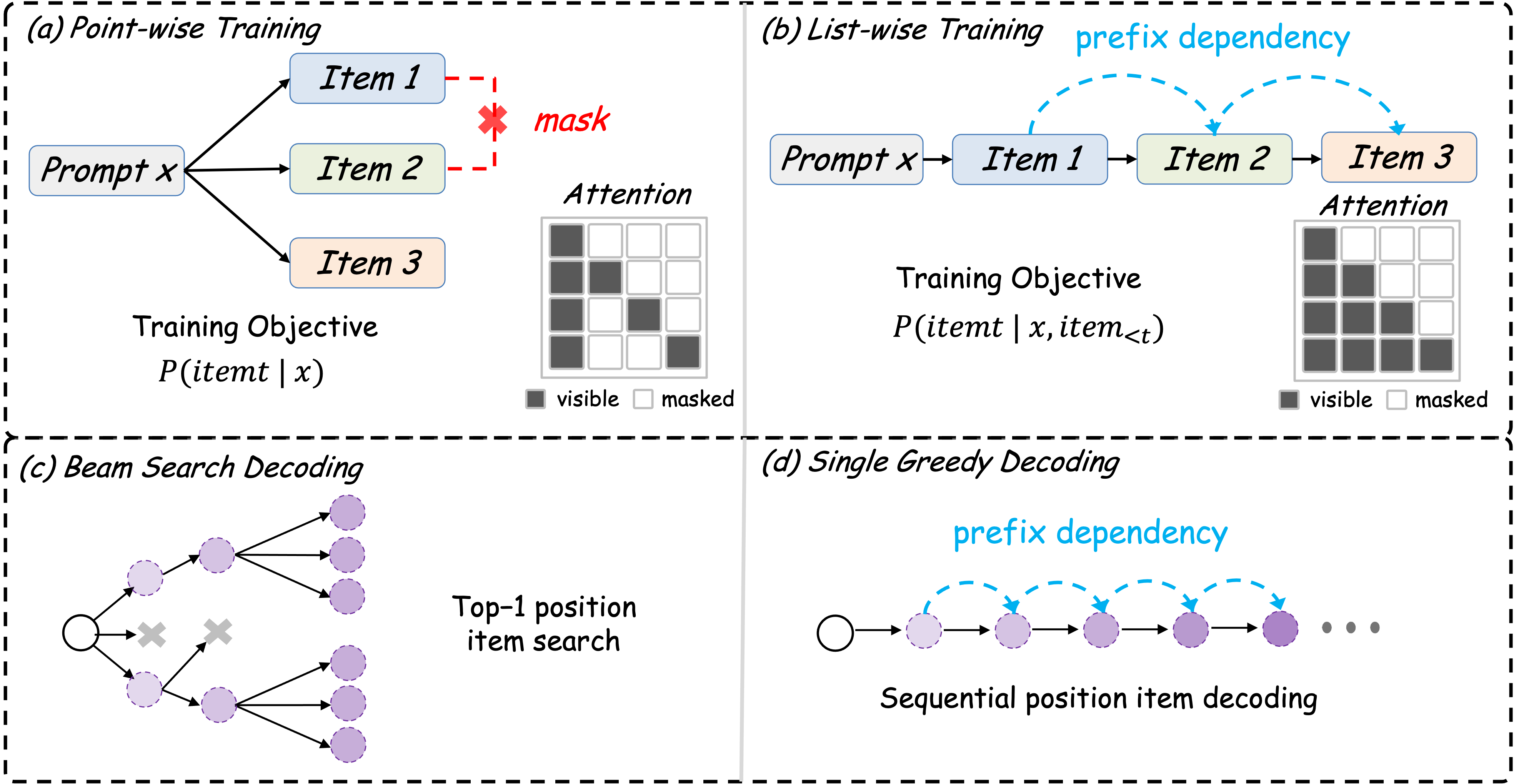}
    \caption{Training and decoding paradigms in SID-based GR.}
    \label{fig:preliminary}
\end{figure}

\subsection{Problem Formulation}
SID-based GR formulates candidate retrieval as autoregressive SID generation. Let $x$ denote the request context, including the query when applicable, user interaction history, and contextual features, and let $\mathcal{I}$ denote the item corpus. Each item $i\in\mathcal{I}$ is represented by a fixed-length SID sequence $\mathbf{s}(i)\in\mathcal{V}^{L}$, where $\mathcal{V}$ is the token vocabulary and $L$ is the number of SID tokens per item. Given $x$, the retrieval model produces an ordered list of $K_d$ candidate items for downstream ranking:
\begin{equation}
\hat{\mathbf{y}}=(\hat{i}_1,\ldots,\hat{i}_{K_d}),\qquad \hat{i}_k\in\mathcal{I}.
\end{equation}

\subsection{Training Formulations}
GR can model individual items or entire lists (\eg, sessions, pages, and slate). We distinguish point-wise and list-wise training according to whether item generation is conditioned on preceding target items in Figure~\ref{fig:preliminary}. In both formulations, the probability of an item is computed by autoregressively factorizing the probability of its SID sequence.

\textbf{Point-wise training} predicts each target item independently of the other target items, conditioned on the request context. For a training example with targets $\mathbf{y}=(i_1,\ldots,i_M)$, the objective aggregates the individual item losses:
\begin{equation}
\mathcal{L}_{\mathrm{point}}(x,\mathbf{y})
=-\sum_{m=1}^{M}w_m\log p_\theta(i_m\mid x),
\end{equation}
where $w_m\geq 0$ is an optional supervision weight based on behavioral feedback, with $w_m=1$ in the unweighted case. Each item is generated without conditioning on other target items.

\textbf{List-wise training} generates the target items sequentially, conditioning each item on the preceding target items. The corresponding objective is
\begin{equation}
\mathcal{L}_{\mathrm{list}}(x,\mathbf{y})
=-\sum_{m=1}^{M}w_m\log p_\theta(i_m\mid x,i_{<m}),
\end{equation}
where $i_{<m}=(i_1,\ldots,i_{m-1})$ denotes the preceding target items, represented by their SID sequences. Both formulations use next-token prediction supervision and the distinction is whether the conditioning context extends across target-item boundaries.

\subsection{Decoding Strategies}
As illustrated in Figure~\ref{fig:preliminary}, item-level beam search and single greedy decoding provide two contrasting strategies for constructing candidate lists at inference.

\textbf{Item-level beam search} maintains up to $B$ active hypotheses, scoring each partial SID sequence by its cumulative log-probability:
\begin{equation}
S_\theta(\mathbf{a}\mid x)
=\sum_{j=1}^{|\mathbf{a}|}\log p_\theta(a_j\mid x,a_{<j}).
\end{equation}
Here, $\mathbf{a}$ is a partial SID sequence, and $a_{<j}$ denotes its prefix before position $j$. After $L$ decoding steps, completed hypotheses are mapped to items, and $K_d$ candidates are selected in beam order. Each search path represents one item rather than a complete list.

\textbf{Single greedy decoding} uses a list-wise model to generate the concatenated SIDs of all $K_d$ items along one autoregressive trajectory:
\begin{equation}
\hat{z}_t=\arg\max_{v\in\mathcal{V}}
p_\theta(v\mid x,\hat{\mathbf{z}}_{<t}),
\qquad t=1,\ldots,K_dL.
\end{equation}
Here, $\hat{\mathbf{z}}_{<t}$ denotes the generated token prefix. Each block of $L$ consecutive tokens represents one item, and later items condition on all preceding generated items. Unlike item-level beam search, single greedy decoding maintains only one evolving prefix and generates the entire candidate list in one continuous decoding pass.

\section{Methodology}
\label{sec:method}

\begin{figure*}
    \centering
    \includegraphics[width=0.98\linewidth]{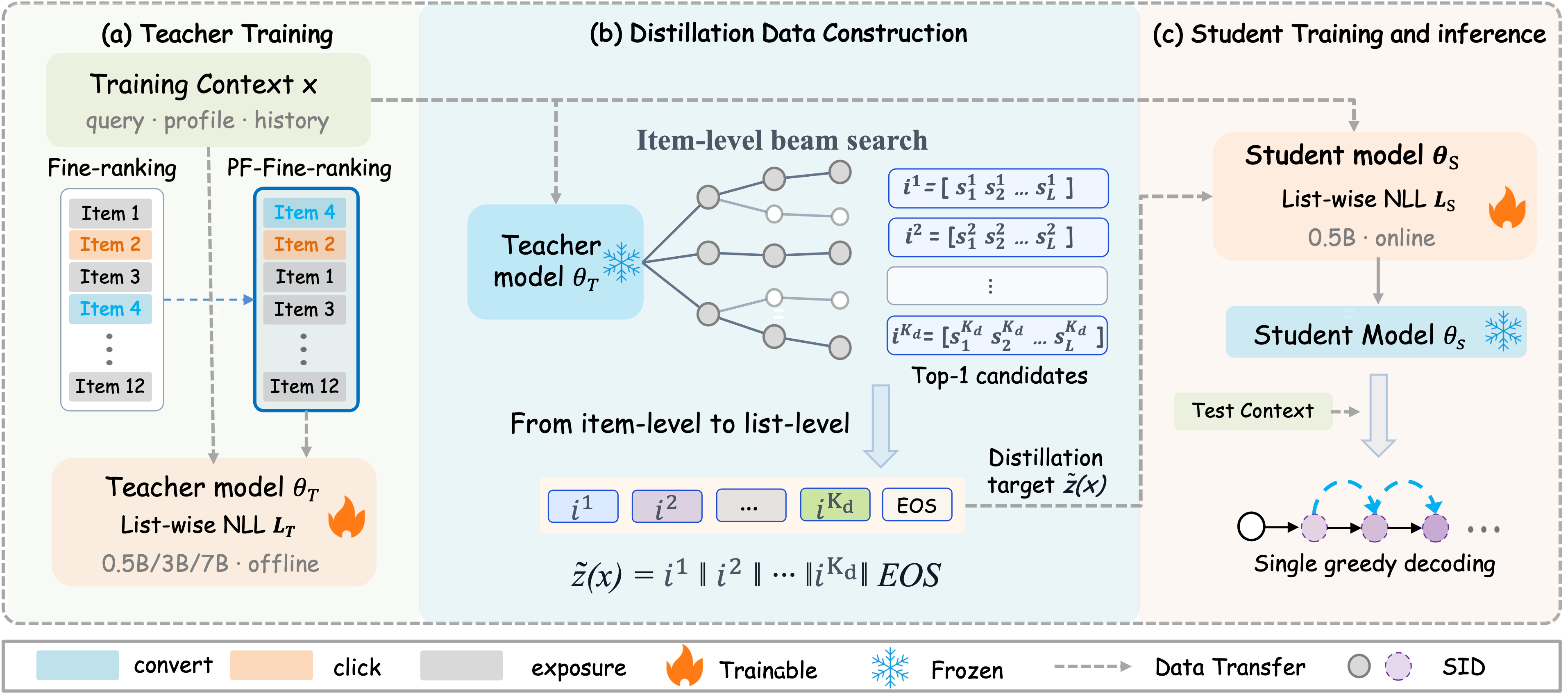}
    \caption{Overview of SLIDE. (a) The teacher is trained on PF-Fine-ranking lists, with converted items first, followed by clicked items and then impressed items. (b) The frozen teacher performs item-level beam search, and the retrieved SIDs are concatenated in descending beam-score order, followed by EOS. (c) A $0.5$B student learns from these list-wise targets and generates a candidate list along a single greedy decoding trajectory.}
    \label{fig:main}
\end{figure*}

\subsection{Overview}
As illustrated in Figure~\ref{fig:main}, SLIDE comprises three stages: teacher training on fine-ranking targets, offline construction of list-wise distillation targets through item-level beam search, and student training for greedy list generation. The key operation is \emph{search-to-list conversion}, which serializes candidates retrieved along separate search paths into successive targets within a single autoregressive trajectory.

\subsection{Teacher Training}

\paragraph{Training formulation.}
Under identical supervision, point-wise and list-wise training yield comparable retrieval performance with item-level beam search (Table~\ref{tab:pointwise_listwise}). We therefore adopt the list-wise formulation introduced in the preliminaries, which also supports direct list generation and allows both decoding strategies to be evaluated using the same model. In this formulation, each target item is conditioned on the preceding target items.

\paragraph{Training targets.}
User feedback is inherently sparse, so exposed but unclicked items from fine-ranking outputs are commonly used as supplementary labels. We construct 12-item teacher targets with converted items first, followed by clicked items and then exposed but unclicked items, preserving the fine-ranking order within each group. We denote this positive-first fine-ranking (PF-Fine-ranking) dataset by
$\mathcal{D}=\{(x_n,\mathbf{y}_n)\}_{n=1}^{N}$,
where $\mathbf{y}_n=(i_{n,1},\ldots,i_{n,12})$.

\paragraph{Training objective.}
Let $\theta_T$ denote the teacher parameters. We assign equal weight to all target items and minimize the list-wise negative log-likelihood:
\begin{equation}
\mathcal{L}_T
=-\frac{1}{N}\sum_{n=1}^{N}\sum_{m=1}^{12}
\log p_{\theta_T}(i_{n,m}\mid x_n,i_{n,<m}).
\end{equation}
Although the teacher is trained on complete lists, subsequent item-level beam search uses only the distribution at the first item position. Thus, retrieved candidates are not conditioned on previously retrieved items.

\paragraph{Model configuration.}
We use Qwen2.5~\cite{yang2024qwen2} teachers with $0.5$B, $3$B, or $7$B parameters. The teachers and student share the same SID representations, vocabulary, and input features, so teacher outputs can be reused directly as student targets. Since the teacher is used only for offline target construction, its capacity can scale without changing the deployed student.

\subsection{Distillation Data Construction}

\paragraph{Item-level search.}
After training, we freeze the teacher and perform item-level beam search on the training request contexts. Let $K_d$ denote the number of candidate items in each distillation target. For each request context $x$, search returns $K_d$ candidate SID sequences:
\begin{equation}
\mathbf{s}^{(k)}
=(s_1^{(k)},\ldots,s_L^{(k)}),
\qquad k=1,\ldots,K_d.
\end{equation}
Each sequence contains $L$ SID tokens representing one candidate item and is scored by its cumulative log-probability:
\begin{equation}
S_T(\mathbf{s}^{(k)}\mid x)
=\sum_{\ell=1}^{L}
\log p_{\theta_T}
(s_\ell^{(k)}\mid x,s_{<\ell}^{(k)}).
\end{equation}
We index the candidates in descending beam-score order, with $\mathbf{s}^{(1)}$ denoting the highest-scoring sequence. Search operates over the SID output space of the teacher rather than being restricted to the observed candidate subset. It can therefore provide additional supervision for items absent from the original interaction labels.

\paragraph{Search-to-list conversion.}
We then concatenate the retrieved SIDs directly in descending beam-score order and append EOS to form a list-wise training target:
\begin{equation}
\tilde{\mathbf{z}}(x)
=\mathbf{s}^{(1)}
\Vert\mathbf{s}^{(2)}
\Vert\cdots\Vert\mathbf{s}^{(K_d)}
\Vert(\mathrm{EOS}),
\end{equation}
where $\Vert$ denotes concatenation. We insert no inter-item separators and perform neither SID-to-item mapping nor validity filtering during target construction. Because each SID has a fixed length, item boundaries are determined by blocks of $L$ tokens. Each target therefore contains $K_dL+1$ tokens.

Finally, we replace the original training labels with these serialized lists while retaining the request contexts:
\begin{equation}
\widetilde{\mathcal{D}}
=\{(x_n,\tilde{\mathbf{z}}(x_n))\}_{n=1}^{N}.
\end{equation}

\subsection{Student Training and Inference}

\paragraph{List-wise distillation.}
We initialize the student from its original pretrained checkpoint rather than from the trained teacher parameters. We then train it on $\widetilde{\mathcal{D}}$ using the serialized lists as hard targets. Let $\theta_S$ denote the student parameters. With teacher forcing, we minimize the uniformly weighted token-level negative log-likelihood:
\begin{equation}
\mathcal{L}_S=-\frac{1}{N}\sum_{n=1}^{N}\sum_{t=1}^{|\tilde{\mathbf{z}}_n|}
\log p_{\theta_S}(\tilde{z}_{n,t}\mid x_n,\tilde{\mathbf{z}}_{n,<t}),
\end{equation}
where $\tilde{\mathbf{z}}_n=\tilde{\mathbf{z}}(x_n)$ includes the terminal EOS token. Training relies solely on offline-generated hard targets and requires neither teacher forward passes nor soft-distribution matching during student optimization.

\paragraph{Single greedy inference.}
At inference, only the student is deployed. Given a test request context $x$, it generates the candidate list through greedy token selection:
\begin{equation}
\hat{z}_t=\arg\max_{v\in\mathcal{V}}
p_{\theta_S}(v\mid x,\hat{\mathbf{z}}_{<t}),
\end{equation}
until EOS is produced. We then split the output into length-$L$ SID sequences, map them to items, and remove invalid and duplicate candidates while preserving their order.

\section{Experiments}
\label{sec:experiments}
We evaluate SLIDE on recommendation and e-commerce search tasks, focusing on retrieval quality under single greedy decoding. We further analyze the effects of training supervision, distillation target construction, teacher and student scaling, and the student optimization budget. Finally, we examine whether list-level RL under single greedy decoding yields larger relative retrieval gains than item-level RL.
\subsection{Experimental Setup}
\label{sec:exp_setup}

\paragraph{Datasets.}
\begin{table}[t]
\centering
\caption{Statistics of datasets}
\label{tab:dataset_statistics}
\normalsize
\setlength{\tabcolsep}{4pt}
\begin{tabular}{lrrr}
\toprule
Dataset & \#Users & \#Items & \#Test \\
\midrule
\textbf{FORGE S3} & 119,916,106 & 255,289,109 & 25,041 \\
\textbf{Taobao Search} & 91,588,112 & 220,716,034 & 16,248 \\
\bottomrule
\end{tabular}
\end{table}

\definecolor{HeaderBg}{HTML}{F1F4F9}
\definecolor{AccentCream}{RGB}{255,255,204}
\definecolor{RowGray}{HTML}{F4F5F8}
\definecolor{RowBlue}{HTML}{EDF2FA}
\definecolor{TextSoft}{HTML}{64748B}
\definecolor{GainRed}{HTML}{DC2626}
\definecolor{DropGreen}{HTML}{15803D}
\begin{table*}[t]
\centering
\caption{Main results on FORGE and Taobao Search. All deployed models have 0.5B parameters. Teacher Size applies only to distilled students (SLIDE). HR@10 changes are relative to Direct Training with Beam Search. Avg counts valid, unique retrieved items. Bold marks the best HR results per dataset.}
\label{tab:main_results}
\setlength{\tabcolsep}{3.5pt}
\renewcommand{\arraystretch}{1.18}
\setlength{\aboverulesep}{0pt}
\setlength{\belowrulesep}{0pt}
\def\DS{\cellcolor{white}}
\def\AC{\cellcolor{AccentCream}}
\def\Best#1{\textbf{#1}}
\def\Ours{\textbf{SLIDE}}
\def\Direct{Direct Training}
\def\Beam{\textcolor{TextSoft}{Beam Search}}
\def\Gain#1{\textcolor{GainRed}{\scriptsize (+#1\%)}}
\def\Drop#1{\textcolor{DropGreen}{\scriptsize (-#1\%)}}
\def\Base{\textcolor{TextSoft}{\scriptsize (ref.)}}
\resizebox{\textwidth}{!}{%
\begin{tabular}{clcclrrlrr}
\toprule
\rowcolor{HeaderBg}
\textbf{Dataset} & \textbf{Method} & \shortstack{\textbf{Student}} & \shortstack{\textbf{Teacher}} & \textbf{Decoding} & \textbf{HR@1} & \textbf{HR@5} & \AC\textbf{HR@10} & \textbf{HR@20} & \shortstack[r]{\textbf{Avg} } \\
\midrule
\rowcolor{RowGray}
\DS & \Direct & 0.5B & -- & \Beam & 0.0067 & 0.0223 & \AC 0.0366\ & 0.0538 & 200.0000 \\
\rowcolor{RowGray}
\DS & \Direct & 0.5B & -- & Greedy & 0.0050 & 0.0050 & \AC 0.0050\,\Drop{86.34} & 0.0050 & 3.9148 \\
\cmidrule(lr){2-10}
\rowcolor{RowBlue}
\DS\textbf{FORGE} & \Ours & 0.5B & 0.5B & Greedy & 0.0064 & 0.0213 & \AC 0.0333\,\Drop{9.02} & 0.0485 & 59.7645 \\
\rowcolor{RowBlue}
\DS & \Ours & 0.5B & 3B & Greedy & 0.0077 & 0.0263 & \AC 0.0397\,\Gain{8.47} & 0.0547 & 68.1863 \\
\rowcolor{RowBlue}
\DS & \Ours & 0.5B & 7B & Greedy & \Best{0.0081} & \Best{0.0292} & \AC\Best{0.0441}\,\Gain{20.49} & \Best{0.0614} & 61.0915 \\
\midrule
\rowcolor{RowGray}
\DS & \Direct & 0.5B & -- & \Beam & 0.1492 & 0.3500 & \AC 0.4498\ & 0.5432 & 195.6491 \\
\rowcolor{RowGray}
\DS & \Direct & 0.5B & -- & Greedy & 0.1318 & 0.2716 & \AC 0.3206\,\Drop{28.72} & 0.3627 & 34.9889 \\
\cmidrule(lr){2-10}
\rowcolor{RowBlue}
\DS\textbf{Taobao Search} & \Ours & 0.5B & 0.5B & Greedy & 0.1470 & 0.3381 & \AC 0.4321\,\Drop{3.94} & 0.5194 & 71.2646 \\
\rowcolor{RowBlue}
\DS & \Ours & 0.5B & 3B & Greedy & 0.1499 & 0.3504 & \AC 0.4469\,\Drop{0.64} & 0.5340 & 69.5151 \\
\rowcolor{RowBlue}
\DS & \Ours & 0.5B & 7B & Greedy & \Best{0.1532} & \Best{0.3628} & \AC\Best{0.4578}\,\Gain{1.78} & \Best{0.5459} & 74.1551 \\
\bottomrule
\end{tabular}%
}
\end{table*}

We conduct experiments on two datasets covering recommendation and e-commerce search. Table~\ref{tab:dataset_statistics} summarizes the dataset statistics.
\begin{itemize}
    \item \textbf{FORGE.} For recommendation, we use the publicly available FORGE dataset~\cite{forge}\footnote{\url{https://huggingface.co/AL-GR}}. Specifically, we use the Stage S3 training and evaluation sets and retain the provided split for reproducibility. Each training sample just contains a single label item.

    \item \textbf{Taobao Search.} For search, we construct an industrial dataset from the Taobao e-commerce search platform. We use requests collected over 10 consecutive days for training and those from the immediately following day for testing to evaluate retrieval on subsequent traffic. Each training sample contains about 12 label items.
\end{itemize}

\paragraph{Baselines.}
Our primary baseline is a directly trained $0.5$B model evaluated with single greedy decoding, matching the model size and inference strategy of the default student. We also report the item-level beam-search performance of the same model as a search-based reference. These comparisons assess whether distillation improves retrieval under single greedy decoding and how the student compares with the beam-search reference.

\paragraph{Evaluation Metrics.}
We report HitRate at $K$ (HR@$K$) for $K\in\{1,5,10,20\}$ to evaluate retrieval quality at small candidate cutoffs. We emphasize \colorbox{AccentCream}{HR@10} because a search session typically contains approximately 10 items. After generation, we map SID sequences to items and remove invalid and duplicate candidates while preserving their order. We then compute HR@$K$ over the first $K$ valid, unique candidates. For the target-ordering analysis, we also report NDCG to assess the ranking quality of the generated lists.

\paragraph{Implementation Details.}
Unless otherwise specified, the student uses Qwen2.5-0.5B~\cite{yang2024qwen2}. For FORGE, we use the SID vocabulary provided by the dataset and retain the single labeled target item in each training instance for direct training. For Taobao Search, we use a three-level SID codebook with sizes $32768\times8192\times8192$. Teacher training uses 12-item fine-ranking lists, with converted items first, followed by clicked items and then impressed items. Within each interaction type, we preserve the original fine-ranking order.
Training is performed on 32 NVIDIA H20 GPUs. SFT uses AdamW with $\beta_1=0.9$ and $\beta_2=0.999$, a global batch size of $8192$, and an initial learning rate of $1\times10^{-4}$ with cosine decay. Teachers are trained for one epoch. Unless otherwise specified, we then construct distillation targets containing $K_d=20$ teacher-generated items per request using a beam-size schedule of $[20,20,20]$. Students are distilled for two epochs to allow additional optimization for greedy retrieval. We examine the effect of this training budget in the epoch ablation in Section~\ref{sec:ablation} and use the final checkpoint for evaluation.
For evaluation on both datasets, item-level beam search uses a beam-size schedule of $[20,20,200]$. Greedy decoding stops when EOS is generated or the maximum decoding length of 1024 tokens is reached.

\subsection{Overall Comparison}
Table~\ref{tab:main_results} compares SLIDE with direct training under single greedy decoding and reports item-level beam search as a reference. The directly trained models exhibit different behavior across the two datasets. On FORGE, greedy decoding achieves low retrieval quality and produces very few candidates. This is consistent with the original single-item targets, which provide no explicit supervision for generating subsequent items and therefore do not train list-level modeling. By contrast, the 12-item fine-ranking targets on Taobao Search support substantially better greedy retrieval. Even with this list-wise supervision, however, greedy decoding still falls well below beam search, with HR@10 of 0.3206 versus 0.4498. Thus, directly training on lists does not by itself close the retrieval gap between the two decoding strategies.

SLIDE addresses this gap by replacing the original targets with serialized teacher beam-search outputs. Even when the teacher and student both have $0.5$B parameters, distillation substantially improves greedy retrieval on both datasets, bringing student performance closer to the beam-search reference. This improvement shows that search-generated targets provide useful supervision beyond the original labels.

Building on this gain, scaling the offline teacher further improves retrieval while keeping the student size fixed. With a $3$B teacher, the student exceeds the directly trained $0.5$B beam-search reference at HR@10 on FORGE and closely matches it on Taobao Search. Increasing the teacher to $7$B enables the student to surpass this reference across all reported HR metrics on both datasets. The benefit of stronger teacher search can therefore be realized through a student without deploying the teacher online.

Alongside these quality gains, the distilled students produce substantially longer candidate lists than the directly trained models under greedy decoding. Together, these results demonstrate that SLIDE enables a $0.5$B student to generate longer, higher-quality lists in a single greedy rollout and benefit from teacher scaling performed entirely offline.

\subsection{Training and Distillation Analysis}
\label{sec:ablation}
SLIDE relies on high-quality teacher search results, effective list-wise distillation targets, and sufficient student learning to translate search quality into single greedy retrieval. The following analyses examine the choices at each stage and explore RL as a further refinement:
\begin{itemize}[leftmargin=*, nosep]
    \item Point-wise versus list-wise training under beam search inference.
    \item Teacher supervision from the behavioral sequence and the fine-ranking sequence.
    \item The number of teacher-searched items used in distillation.
    \item The contributions of teacher-searched items set and their order.
    \item The effects of model scaling and the optimization budget.
    \item Gains from list-level RL under single greedy decoding versus item-level RL under beam search.
\end{itemize}

\begin{table}[t]
\centering
\caption{Point-wise and list-wise teachers achieve comparable Beam Search retrieval quality under the same labels.}
\label{tab:pointwise_listwise}
\small
\setlength{\tabcolsep}{5pt}
\renewcommand{\arraystretch}{1.10}
\begin{tabular}{llrr>{\columncolor{AccentCream}}rr}
\toprule
Training & Weighting & HR@1 & HR@5 & HR@10 & HR@20 \\
\midrule
Point-wise & \texttimes & 0.1093 & 0.2882 & 0.3911 & 0.4917 \\
List-wise & \texttimes & \textbf{0.1126} & \textbf{0.2916} & \textbf{0.3934 }& \textbf{0.4978} \\
\midrule
Point-wise & \checkmark & 0.1352 & 0.3173 & \textbf{0.4125} & \textbf{0.5046} \\
List-wise & \checkmark & \textbf{0.1383} & \textbf{0.3183} & 0.4078 & 0.5007 \\
\bottomrule
\end{tabular}
\end{table}

\paragraph{\textbf{Point-wise vs.\ List-wise Training.}}
Our main comparison uses a directly trained list-wise model as the baseline. Before investigating how to improve teacher quality, we first examine whether this training formulation limits its beam-search performance. Since item-level beam search retrieves each candidate along a separate SID trajectory, point-wise training is a natural alternative and may provide a stronger search reference. This comparison helps determine whether the gains in Table~\ref{tab:main_results} depend on the choice of direct-training baseline.
To isolate the effect of the training formulation, we train point-wise and list-wise teachers using the same supervision labels and evaluate them under the same beam-search configuration. We consider both unweighted and label-weighted~\cite{rankgr} training to assess whether the comparison is consistent across weighting settings. As shown in Table~\ref{tab:pointwise_listwise}, without weighting, list-wise training performs slightly better across all reported metrics, with an HR@10 advantage of only 0.23pt. With weighting, list-wise training achieves higher HR@1 and HR@5, while point-wise training has a small advantage at HR@10 and HR@20. Overall, the two training formulations show no substantial differences in retrieval performance.
These results suggest that using a list-wise teacher does not substantially weaken the beam-search reference relative to a point-wise alternative. The comparison therefore supports the use of a directly trained list-wise model as the baseline in Table~\ref{tab:main_results}. Given their comparable search quality, we retain list-wise training in our experiments. This choice allows the same model to support both item-level beam search and direct list generation, while keeping the training formulation unchanged when the original labels are replaced with serialized teacher-search outputs.

\begin{table}[t]
\centering
\caption{Effect of supervision sources on teacher retrieval quality under Beam Search.}
\label{tab:supervision_sources}
\small
\setlength{\tabcolsep}{5pt}
\renewcommand{\arraystretch}{1.10}
\begin{tabular}{lrr>{\columncolor{AccentCream}}rr}
\toprule
Supervision Source & HR@1 & HR@5 & HR@10 & HR@20 \\
\midrule
\textsc{Behavior} & 0.1126 & 0.2916 & 0.3934 & 0.4978 \\
\textsc{Fine-Rank} & 0.1458 & 0.3386 & 0.4384 & 0.5337 \\
\textsc{PF-Fine-Rank} & \textbf{0.1492} & \textbf{0.3500} & \textbf{0.4498} & \textbf{0.5432} \\
\bottomrule
\end{tabular}
\end{table}

\paragraph{\textbf{Effect of Supervision Sources.}}
Having established that point-wise and list-wise training have comparable beam-search quality, we next examine how to obtain a stronger teacher. This is important for SLIDE because the teacher's search results determine the quality of the offline distillation targets. We compare conventional behavioral supervision (\textsc{Behavior}) with lists produced by the fine-ranking model (\textsc{Fine-Rank}), which provide richer targets beyond observed interactions.
We further construct \textsc{PF-Fine-Rank} by moving the positive item to the top position while preserving the relative order of the remaining items. Under list-wise training, placing the positive item first gives it direct supervision under the initial context used by item-level beam search.
Table~\ref{tab:supervision_sources} shows that \textsc{Fine-Rank} outperforms \textsc{Behavior} across all reported metrics, increasing HR@10 from 0.3934 to 0.4384. Building on this improvement, \textsc{PF-Fine-Rank} provides further gains across all cutoffs. These results suggest that fine-ranking targets improve teacher retrieval quality, and that prioritizing observed positives makes this supervision more effective for item-level beam search. We therefore adopt \textsc{PF-Fine-Rank} for teacher training in experiments.

\begin{table}[t]
\centering
\caption{Effect of distillation target length on student retrieval under single greedy decoding with a fixed teacher.}
\label{tab:teacher_output_length}
\small
\setlength{\tabcolsep}{5pt}
\renewcommand{\arraystretch}{1.10}
\begin{tabular}{crrrrr}
\toprule
Length & HR@1 & HR@12 & HR@20 & HR@50 & Avg.\ \#Candidates \\
\midrule
12 & \textbf{0.1418} & \textbf{0.4387} & 0.4885 & 0.5349 & 46.4 \\
20 & 0.1417 & 0.4361 & \textbf{0.4937} & 0.5609 & 72.7 \\
50 & 0.1394 & 0.4262 & 0.4841 & \textbf{0.5698} & 123.4 \\
\bottomrule
\end{tabular}
\end{table}

\paragraph{\textbf{Teacher Output Length.}}
After establishing the teacher training setup, we examine how many search-generated items should be included in each distillation target. Constructing longer targets substantially increases the resource cost of offline distillation data generation, making it important to determine whether the additional items improve retrieval at the intended cutoff. We therefore vary the target length while keeping the teacher fixed and evaluate all students using single greedy decoding.
Table~\ref{tab:teacher_output_length} shows that the benefit of longer targets depends on the evaluation depth. The 12-item targets achieve the highest HR@12, while increasing the length to 20 yields the best HR@20. Further extending the targets to 50 produces the highest HR@50 and substantially more retrieved candidates, but does not improve the shallower-cutoff metrics. Thus, longer targets improve deeper retrieval coverage rather than providing uniform gains across cutoffs.
Across the tested settings, each target length achieves the best result at its corresponding cutoff. This suggests that when HR@$K$ is the primary objective, using $K_d=K$ teacher-generated items per target is sufficient: extending the target beyond $K$ adds offline construction cost without improving the target metric in our experiments. Although we primarily focus on HR@10, we use $K_d=20$ in the main experiments to support a broader evaluation.

\begin{figure}
    \centering
    \includegraphics[width=0.98\linewidth]{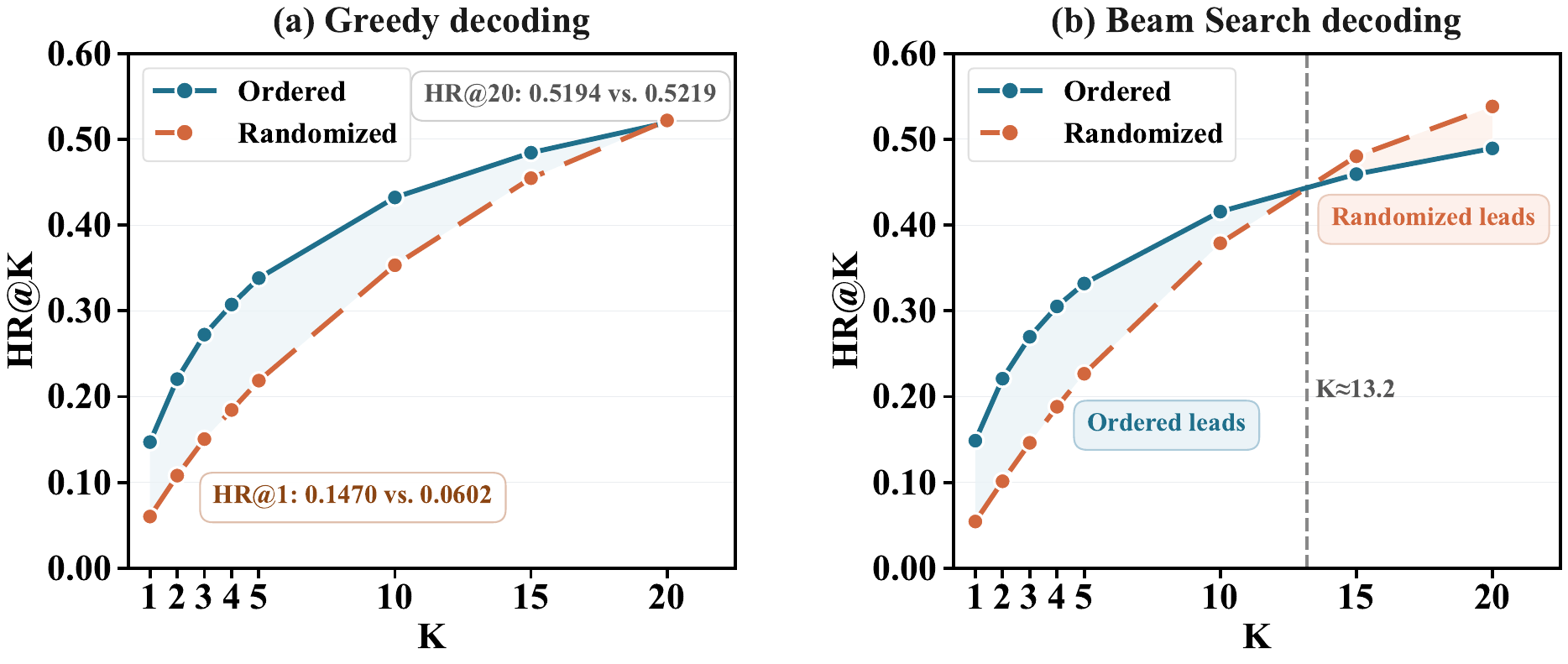}
    \caption{Effect of distillation target ordering.}
    \label{fig:item_ordering}
\end{figure}

\begin{figure*}
    \centering
    \includegraphics[width=0.98\linewidth]{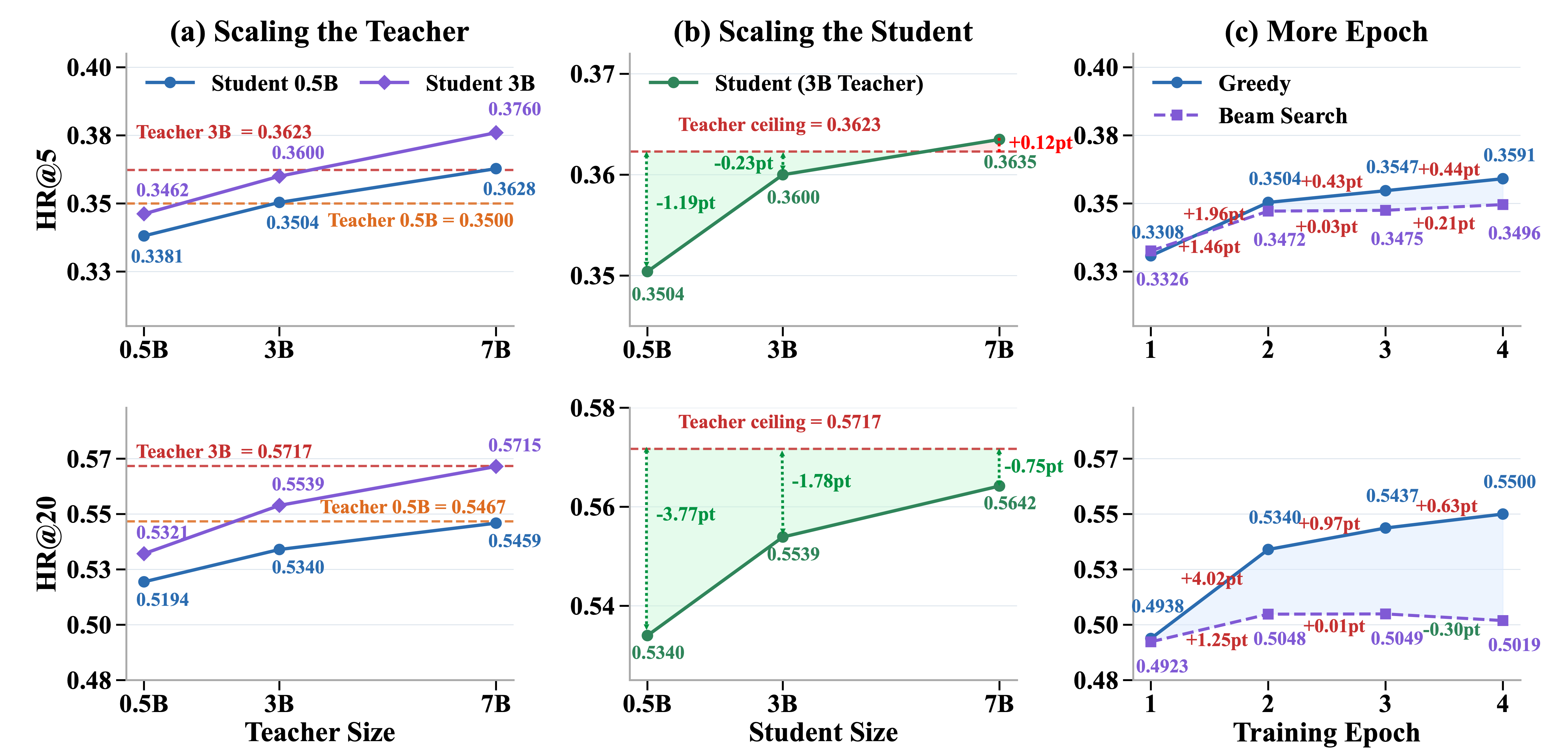}
    \caption{Scaling analysis of model and optimization budget.}
    \label{fig:scaling}
\end{figure*}
\paragraph{\textbf{Distillation Target Ordering.}}
Having examined target length, we next investigate how the teacher-searched items should be ordered for distillation. To isolate the contribution of ordering, we randomly permute the items within each target while preserving the item set. Figure~\ref{fig:item_ordering} shows that randomization substantially degrades greedy head retrieval, while HR@20 remains nearly unchanged. These results suggest that the distilled item set supports candidate quality and coverage, whereas its ordering guides the order in which the student retrieves those candidates. Consistent with this distinction, NDCG@20 declines from 0.3010 to 0.2271, showing that only the same candidate set does not ensure comparable ranking quality.
Beyond these changes in greedy retrieval, randomization also reverses the relative performance of beam search and greedy decoding at deeper cutoffs: beam search underperforms greedy decoding with ordered targets but outperforms it with randomized targets. We hypothesize that randomization distributes first-position supervision across more candidates, potentially flattening the first-item prediction distribution and making lower-ranked items more accessible to item-level beam search. Overall, preserving teacher ordering helps the student translate the distilled candidate set into a well-ranked list through a single greedy rollout.

\paragraph{\textbf{Scaling Teachers and Students.}}
We next examine how teacher size, student size, and additional optimization affect SLIDE. Figure~\ref{fig:scaling} compares these three scaling directions.

\noindent \textbf{(1) Larger teachers raise attainable student performance.}
With student size fixed, scaling the teacher consistently improves retrieval in Figure~\ref{fig:scaling}(a). More importantly, these gains can extend beyond the corresponding direct-training baseline: the 3B student distilled from a 7B teacher achieves higher HR@5 than the directly trained 3B model (0.3760 vs.\ 0.3623), while matching its HR@20. Stronger teacher-search supervision therefore raises the performance attainable by a fixed-size student without increasing student-side inference cost.

\noindent \textbf{(2) Larger students learn more effectively.}
Beyond supervision quality, we examine the role of student capacity. Figure~\ref{fig:scaling}(b) fixes the 3B teacher and its distillation data while varying student size. The HR@20 gap to the teacher decreases from 3.77pt for the 0.5B student to 0.75pt for the 7B student, whose HR@5 also slightly exceeds the teacher's beam-search result. Since the supervision remains unchanged, the narrowing gap suggests that students with greater capacity can more effectively absorb the knowledge provided by the teacher through distillation.

\noindent \textbf{(3) Head retrieval transfers more readily.}
The capacity gap also varies with retrieval depth. With a 7B teacher, the 0.5B student reaches the HR@5 level of the directly trained 3B model (0.3628 vs.\ 0.3623), but remains below it at HR@20 (0.5459 vs.\ 0.5717). Increasing the student to 3B nearly closes this deeper-cutoff gap. The easier transfer of head retrieval reflect the already smaller performance gap between smaller and larger models at early ranks.

\noindent \textbf{(4) Scaling training epochs offers further gains.}
While both model-scaling directions improve retrieval, larger teachers increase the costs of teacher training and distillation data construction, and student size is often constrained by online inference latency. Increasing distillation data is therefore a practical alternative. We examine epoch scaling, which reuses the existing data rather than requiring additional distillation data.
Figure~\ref{fig:scaling}(c) shows that both decoding strategies improve substantially between the first two epochs, but their subsequent trends diverge. Greedy retrieval continues to improve, gaining another 1.60pt at HR@20 from epoch two to four, whereas beam-search performance largely saturates. A possible explanation is that learning to generate an entire ordered list is more demanding than learning the first-item predictions used by item-level beam search. Repeated training on the same targets can therefore continue to benefit greedy list generation, although the smaller increments indicate diminishing returns.

\begin{table}[t]
\centering
\small
\caption{RL refinement under different decoding schemes. \textit{Gain} denotes the relative improvement over SFT.}
\label{tab:rl_greedy_vs_beam}
\begin{tabular*}{\columnwidth}{@{\extracolsep{\fill}}llcc>{\columncolor{AccentCream}}cc@{}}
\toprule
\textbf{Decoding} & \textbf{Method} & \textbf{HR@1} & \textbf{HR@5} & \textbf{HR@10} & \textbf{HR@20} \\
\midrule
\multirow{3}{*}{Item-level} & SFT & 0.1557 & 0.3390 & 0.4257 & 0.5098 \\
& RL & 0.1570 & 0.3410 & 0.4307 & 0.5132 \\
& \textbf{\textit{Gain}} & \textcolor{GainRed}{\textbf{+0.84\%}} & \textcolor{GainRed}{\textbf{+0.59\%}} & \textcolor{GainRed}{\textbf{+1.17\%}} & \textcolor{GainRed}{\textbf{+0.67\%}} \\
\midrule
\multirow{3}{*}{List-level} & SFT & 0.1916 & 0.4120 & 0.5061 & 0.5868 \\
& RL & 0.2063 & 0.4317 & 0.5267 & 0.6071 \\
& \textbf{\textit{Gain}} & \textcolor{GainRed}{\textbf{+7.67\%}} & \textcolor{GainRed}{\textbf{+4.78\%}} & \textcolor{GainRed}{\textbf{+4.07\%}} & \textcolor{GainRed}{\textbf{+3.46\%}} \\
\bottomrule
\end{tabular*}
\end{table}

\paragraph{\textbf{RL for List-Level Generation and Item-Level Retrieval}}
As a supplementary validation of our motivation, we compare RL refinement for list-level generation and item-level retrieval. Both settings use sampled rollouts during training. For list-level generation, each rollout produces a complete list, allowing retrieval decisions at different positions to be refined within a shared autoregressive trajectory rather than making individual candidates compete solely for the top-1 position. Both settings use item-level efficiency rewards, with an additional repetition penalty for list-level generation. Table~\ref{tab:rl_greedy_vs_beam} shows that RL for list-level generation yields relative improvements of 3.46--7.67\% across cutoffs, compared with 0.59--1.17\% for item-level retrieval. These results support the effectiveness of list-level generation for subsequent RL refinement in our setting. Implementation details and relevance gains are provided in Appendix~\ref{app:rl}.

\section{Conclusion}
We presented \textbf{SLIDE}, a search-to-list distillation framework that transfers offline item-level beam-search outputs into single-rollout generative retrieval. By serializing teacher-retrieved SIDs in beam-score order, SLIDE converts independently searched candidates into sequential list-wise supervision, enabling the student to generate the retrieval list through a single greedy pass. Experiments on FORGE and Taobao Search show substantial improvements over directly trained greedy decoding and, with stronger offline teachers, performance surpassing the beam-search reference at the same deployment size. Our analyses highlight the complementary roles of teacher item selection and ordering, as well as the benefits of scaling model capacity and student optimization budget. Supplementary experiments also show larger RL gains for list-level generation than for item-level. Together, these findings establish offline search-to-list distillation as a practical way to improve single-rollout retrieval while decoupling teacher scaling from online deployment.

\section{Limitations}
SLIDE aims to provide a strong foundation for direct list-wise generation, facilitating extensions such as CoT or RL and the optimization of list-level objectives such as relevance and diversity. However, its scope is limited to end-to-end list generation rather than large-scale candidate generation for downstream ranking. Constructing the distillation data incurs substantial offline computational costs, as it requires teacher beam search over training contexts. Moreover, SLIDE involves a three-stage pipeline, which adds complexity and makes model iteration more difficult. A valuable direction for future work is to enable single-pass greedy decoding to match beam-search retrieval quality through a single training stage, reducing both pipeline complexity and the cost of model iteration.

\bibliographystyle{ACM-Reference-Format}
\bibliography{main}

\appendix
\section{Appendix}

\subsection{Use of Large Language Models}
We use the publicly available Qwen model as the backbone for our experiments. For writing, generative AI tools are used solely to improve clarity, coherence, and presentation. All figures, initial drafts, and substantive content are independently prepared by the authors.

\subsection{RL Details}
\label{app:rl}

We describe efficiency-oriented RL for two generative retrieval paradigms: \emph{item-level retrieval} with item-level beam search and \emph{list-level generation} with single-pass greedy decoding. Each item is represented by a triplet of Semantic IDs (SIDs). In item-level retrieval, each response represents one item, whereas in list-level generation, each response represents an ordered list produced within a single autoregressive trajectory. Both settings use the efficiency reward
\[
r = \mathrm{CTR} \times \mathrm{CVR},
\]
provided by an efficiency scoring service. List-level generation additionally applies a repetition penalty.

\subsubsection{RL for Item-Level Retrieval}

For each input, we sample $n$ responses during rollout, each containing the SID triplet of one item. These triplets are mapped to their corresponding items and scored by the efficiency service. We normalize the rewards within the sampled group to obtain relative advantages, assign each advantage to its corresponding response, and update the policy using standard GRPO.
Because each response contains a single item, response-level and item-level rewards have the same granularity. Candidates compete within the sampled group for higher relative advantages rather than being jointly optimized as a complete list. The training candidate group is formed through sampling, whereas the deployed Top-$K$ list is assembled through item-level beam search.

\subsubsection{RL for List-Level Generation}

For each input, we sample $n$ responses during rollout, each containing an ordered list whose length may vary. We extract the SID triplets from all sampled responses and deduplicate them to form a candidate set for that input. Each unique triplet is then mapped to its corresponding item and assigned an efficiency score.
Within this candidate set, items with valid efficiency scores receive Z-score-normalized advantages:
\[
A(s) = \frac{r(s) - \mu}{\sigma + \epsilon},
\]
where $s$ denotes an item represented by a SID triplet, $\mu$ and $\sigma$ are the mean and standard deviation of valid rewards in the candidate set, and $\epsilon$ ensures numerical stability.

We assign $A(s)$ to the tokens encoding item $s$, rather than sharing one advantage across the entire response. These token-level advantages are used in a GRPO-based policy update, without further response-level normalization of the efficiency signal. Items above and below the candidate-set mean can therefore receive positive and negative advantages within the same response. This enables fine-grained credit assignment across list positions while retaining the full list as a single rollout.

We additionally apply a repetition penalty to discourage duplicate items and prevent repeated generation of high-reward items. For each repeated occurrence, the tokens encoding the item receive no efficiency advantage, and the final token receives a negative advantage. At inference, the refined model generates the full Top-$K$ list through single-pass greedy decoding.

\subsection{Additional Experiments}

\subsubsection{Relevance Optimization through RL}

A key motivation for moving from beam search to list-wise modeling is to better optimize search relevance and improve user experience. Beyond the HR metrics reported in the main text, we therefore evaluate relevance-oriented RL after SLIDE. Relevance at each cutoff is assessed using a model developed by our internal relevance team.
As shown in Table~\ref{tab:rl_relevance}, RL with a relevance reward improves scores by 2.69--3.22 percentage points over the SLIDE baseline across the reported cutoffs. Further incorporating an LLM-as-a-judge assessment of self-consistency into the reward yields additional gains at every cutoff. These results demonstrate that SLIDE provides a foundation for extending list-wise RL beyond efficiency optimization to relevance-oriented objectives.

\begin{table}[!tbp]
\centering
\caption{Relevance optimization through RL after SLIDE.}
\label{tab:rl_relevance}
\small
\setlength{\tabcolsep}{4pt}
\renewcommand{\arraystretch}{1.10}
\resizebox{\columnwidth}{!}{%
\begin{tabular}{lrr>{\columncolor{AccentCream}}rr}
\toprule
Method & Rel.@1 & Rel.@5 & Rel.@10 & Rel.@20 \\
\midrule
SLIDE baseline
& 0.8637 & 0.8510 & 0.8422 & 0.8352 \\
+ Relevance RL
& 0.8906 & 0.8814 & 0.8744 & 0.8674 \\
+ LLM-as-a-judge self-consistency
& \textbf{0.8951} & \textbf{0.8865}
& \textbf{0.8792} & \textbf{0.8716} \\
\bottomrule
\end{tabular}%
}
\end{table}

\begin{table}[!tbp]
\centering
\caption{Effect of conversion-first distillation targets. Both teacher and student have 0.5B parameters.}
\label{tab:conversion_reintroduction}
\small
\setlength{\tabcolsep}{4pt}
\renewcommand{\arraystretch}{1.10}
\resizebox{\columnwidth}{!}{%
\begin{tabular}{llrr>{\columncolor{AccentCream}}rr}
\toprule
Target & Decoding & HR@1 & HR@5 & HR@10 & HR@20 \\
\midrule
Original & Greedy
& \textbf{0.1385} & \textbf{0.3205}
& \textbf{0.4030} & \textbf{0.4698} \\
Conversion-first & Greedy
& 0.1312 & 0.3201 & 0.4018 & 0.4682 \\
\midrule
Original & Beam Search
& 0.1395 & 0.3152 & 0.3867 & 0.4576 \\
Conversion-first & Beam Search
& \textbf{0.1414} & \textbf{0.3318}
& \textbf{0.4229} & \textbf{0.5067} \\
\bottomrule
\end{tabular}%
}
\end{table}

\subsubsection{Reintroducing Conversions into Distillation Targets}

SLIDE constructs distillation targets through teacher beam search rather than directly from the original supervision labels. Only approximately 60\% of these targets contain the ground-truth converted item. We therefore investigate whether explicitly prioritizing this item benefits the distilled student. In this conversion-first variant, we move the item to the first position if it is already present, or insert it at the beginning otherwise. In both cases, we preserve the relative order of the remaining items.
As shown in Table~\ref{tab:conversion_reintroduction}, this modification does not improve single-pass greedy decoding, with slight decreases across the reported HR metrics. In contrast, it consistently improves item-level beam search, including gains of 3.62pt and 4.91pt in HR@10 and HR@20, respectively. One possible explanation is that placing the converted item first reinforces its supervision under the initial decoding context, which is directly relevant to item-level beam search. These results suggest that prioritizing the original conversion supervision is more beneficial for the student's beam-search capability than for direct list generation. Note that the results are obtained before training has converged. They therefore reflect an intermediate-stage comparison rather than final converged performance.

\end{document}